\documentclass{ifacconf}

\counterwithin*{section}{part}

\usepackage{graphicx}      
\usepackage{natbib}        
\usepackage{amsmath}
\usepackage{amssymb}

\usepackage[capitalise]{cleveref}

\begin{document}
\begin{frontmatter}

\title{HAVOK Model Predictive Control for Time-Delay Systems with Applications to District Heating} 

\author[First]{Christian M. Jensen} 
\author[First]{Mathias C. Frederiksen} 
\author[First,Second]{Carsten S. Kallesøe}
\author[First]{Jeppe N. Jensen}
\author[First]{Laurits H. Andersen}
\author[First,Third]{Roozbeh Izadi-Zamanabadi}

\address[First]{Department of Electronic Systems, Automation and Control, Aalborg University, 9210 Aalborg, Denmark (e-mail: cmje21@student.aau.dk)}
\address[Second]{Grundfos A/S, Poul Due Jensens Vej, Bjerringbro, Denmark (e-mail: ckallesoe@grundfos.com)}
\address[Third]{Danfoss A/S, Nordborgvej 81, Nordborg, Denmark (email: roozbeh@danfoss.com)}

\begin{abstract}                
A computationally efficient Model-Predictive Control (MPC) approach is proposed for systems with unknown delay using only input/output data. We use the Koopman operator framework and the related Hankel Alternative View of Koopman (HAVOK) algorithm to identify a model in a basis of projected time-delay coordinates and demonstrate a novel MPC structure that reduces and bounds the computational complexity. The proposed HAVOK-MPC approach is validated experimentally on a laboratory-scale District Heating System (DHS), demonstrating excellent prediction and tracking performance while only requiring knowledge of a conservative upper bound on the system delay.
\end{abstract}


\end{frontmatter}
\section{Introduction}\label{sec:Intro}
District heating systems (DHS) are a method for supplying centrally generated heat to residential and commercial consumers. They are common in the Nordics and the Baltic, and of significant interest in the context of energy decarbonization because they allow re-use of industrial waste heat as well as heat from a variety of power generation sources such as Combined Heat and Power (CHP) plants and Waste-to-Energy (WtE) plants, see e.g. \cite{Werner:2017}.

The control problem in any DHS is to deliver a desired temperature at the consumers by regulating the supply temperature at the central plant. However, as heat is distributed via a fluid medium, DHS have transport delays that depend on flow rate, distance, and pipe diameter. While time-delayed systems in general are well-studied - see e.g. \cite{Gu:2003} - this is less so the case for systems such as DHS where delays are input-dependent. Recent work on this latter class of systems is based on a Lyapunov-Krasovskii (LK) functional approach as in the references \cite{Bresch:2014,Bendtsen:2017}, the latter of which applies the approach directly to a small-scale DHS.

However, the LK approach is inherently model-based, and as seen in e.g. \cite{Gabrielaitiene:2007}, DHS are often complex networks with many potential variables of interest and unknown parameters, making first-principles modelling daunting, and input-output data-driven approaches attractive. A promising tool for the latter is the Koopman operator framework, first defined in \cite{Koopman:1931} and with a recent resurgence in interest due to the work of \cite{Mezic:2004}. This framework describes dynamical systems, whether linear or non-linear, in terms of the evolution of their observables under a linear operator, and is connected to the powerful Dynamic Mode Decomposition (DMD) family of data-driven algorithms first proposed in \cite{Schmid:2010}. Critically, a common variant of DMD is taken over time-delayed measurements, inherently connecting it to time-delayed systems, as recently exploited by \cite{Louw:2022}.

This paper proposes a data-driven Model Predictive Control (MPC) framework for time-delay systems. We utilize the Hankel Alternative View of Koopman (HAVOK) algorithm of \cite{Brunton:2016} to propose a novel MPC structure that minimizes the dimensionality issues normally induced by the use of delay coordinates. We then demonstrate the practical feasibility of the approach by applying our proposed approach to a small-scale DHS.

The rest of the paper is organised as follows: in \cref{sec:Prelim} a brief introduction to Koopman operator theory and the HAVOK algorithm is provided. \Cref{sec:HAVOKMPC} details the proposed HAVOK-MPC approach, including a causality-enforcing sparse quadratic programme (QP) structure. The dynamics of a DHS and the physics-informed argument for the use delay coordinates are then introduced in \cref{sec:DistrictHeating}, whereafter \cref{sec:LabResults} presents the experimental results alongside brief discussion. Finally, \cref{sec:Conclusion} summarizes the contributions of the work.

\section{Koopman Operator Theory}\label{sec:Prelim}
In this section, we give a brief summary of the state-of-art pertaining to the Koopman operator framework and the HAVOK algorithm. For an arbitrary discrete-time dynamical system of the form:

\begin{equation}\label{eq:AutoSS}
	x_{k+1} = f(x_k)
\end{equation}

observables $g(x)$ evolve in time under the Koopman operator $\mathcal{K}$ (\cite{Mezic:2005}): 

\begin{equation}\label{eq:Koopman}
	\mathcal{K} \circ g = g(f(x_k)) = g(x_{k+1})
\end{equation}

Critically, this forward mapping in time is linear under the Koopman operator regardless of the underlying dynamical system. For practical data-driven purposes, the Koopman operator is approximated via the Dynamic Mode Decomposition family of algorithms. The basic form of the algorithm is, given two sequential-in-time data matrices $X_k$ and $X_{k+1}$:

\begin{equation}\label{eq:DMD}
	\displaystyle{\min_{A}} \quad X_{k+1} - AX_{k}
\end{equation} 

such that $A \approx \mathcal{K}$ on the subspace spanned by the observables. The algorithm defined in \cref{eq:DMD}, however, often suffers from a lack of linear consistency - as defined in \cite{Tu:2014} - when applied directly to experimental data, and on account of this the observable space is often augmented artificially. This can be achieved via a function library as in \cite{Williams:2015}, by time-delayed coordinates as in \cite{Arbabi:2017} (connecting DMD to the seminal work of \cite{Takens:1981}), or a hybrid combination of the two as in \cite{Korda:2018}. Additionally, the theory of Koopman operators was extended to non-autonomous systems by \cite{Mezic:2016}, and DMD extended in similar fashion by \cite{Proctor:2016}, taking the form known as DMDc:

\begin{equation}\label{eq:DMDc}
	\displaystyle{\min_{G}} \quad X_{k+1} - G\Omega = X_{k+1} - \begin{bmatrix}A & B\end{bmatrix}\begin{bmatrix}X_k \\ \mathcal{U}_k\end{bmatrix} 
\end{equation}

where $\mathcal{U}_k$ is the forcing matrix associated with $X_k$, and $G \approx \mathcal{K}$.

The DMD variant of interest in this paper - the Hankel Alternative View of Koopman (HAVOK) algorithm identified in \cite{Brunton:2016} - belongs to the latter hybrid category. It performs DMD on time-delay coordinates projected onto an orthogonal basis of Singular Value Decomposition (SVD) modes, a basis that has been shown (\cite{Arbabi:2017,Dylewsky:2022}) to have very favourable properties with respect to Koopman operator approximations. This basis is obtained from a matrix of embedded system trajectories that will have either Hankel or Toeplitz structure depending on whether a forwards or backwards (see \cite{Parlitz:2014}) embedding is chosen; an example with Toeplitz structure can be seen in \cref{eq:ToeplitzMat}.

\begin{equation}\label{eq:ToeplitzMat}
	\begin{gathered}
			H = 
			\begin{bmatrix}
				x(k) & x(k+1)  & \hdots & x(k+m) \\ 
				x(k-1) & x(k)  & \hdots & x(k+m-1) \\
				\vdots & \vdots & \vdots & \vdots \\ 
				x(k-d) & x(k+1-d) & \hdots & x(k+m-d)
			\end{bmatrix}
		\\ x \in \mathbb{R}^{n \times 1}, \quad H \in \mathbb{R}^{n\cdot (d+1) \times (m+1)}
	\end{gathered}
\end{equation} 

Here we assume $x(k)$ is directly measurable, such that each column of $H$ can be considered a state vector of a delay-embedded system. $H$ can then be interpreted in terms of repeated composition with the Koopman operator. Let $\delta = k-d$, then:

\begin{equation}\label{eq:ToeplitzMatKoop}
	H = 
	\begin{bmatrix}
		\mathcal{K}^{d}x(\delta) & \mathcal{K}^{d+1}x(\delta)  & \hdots & \mathcal{K}^{d+m}x(\delta) \\ 
		\mathcal{K}^{d-1}x(\delta) & \mathcal{K}^{d}x(\delta)  & \hdots & \mathcal{K}^{d+m-1}x(\delta) \\
		\vdots & \vdots & \vdots & \vdots \\ 
		x(\delta) & \mathcal{K}x(\delta) & \hdots & \mathcal{K}^{m}x(\delta)
	\end{bmatrix}
\end{equation} 

The HAVOK algorithm takes the SVD of \cref{eq:ToeplitzMat} such that:

\begin{equation}\label{eq:ToeplitzSVD}
	H = USV^T
\end{equation} 

where $V^T$ contains principal component trajectories (PCTs) of $H$. Let $H_{k:m-1}$ denote a submatrix of $H$ containing all its rows and the $k$th to $(m-1)$th columns. The principal component trajectories can then be recovered by splitting $H$ and projecting the submatrices onto the PCT space as $V^T_k = S^{-1}U^TH_{k:m-1}$ and $V^T_{k+1} = S^{-1}U^TH_{k+1:m}$. Applying DMD to recover a Koopman approximation on the PCTs then constitutes HAVOK:

\begin{equation}\label{eq:HAVOK}
	\displaystyle{\min_{A_v}} \quad V^T_{k+1} - A_vV^T_{k}, \quad A_v \approx \mathcal{K}, \quad V_i^T \in \mathbb{R}^{r\times m-1}
\end{equation} 

where $r$ is the dimension of the PCT space, which is often truncated. For non-autonomous systems, forcing can be projected onto the PCTs after forming an embedded input matrix $H_u$ analogous to $H$, and \cref{eq:DMDc} applied as per usual, resulting in $G = [A_v \ B_v] \approx \mathcal{K}$.

%
%

\section{Causality-Enforcing HAVOK-MPC}\label{sec:HAVOKMPC}
 In the following section, we assume a linear model of the system dynamics has already been found via the HAVOK approach described by \cref{eq:HAVOK}.
 
 A fundamental issue posed by the use of delay embedding models with MPC is that the former often requires a large state dimension to reconstruct dynamics. For example, \cite{Champion:2019} use embedding lengths in excess of $100$, and while a condensed QP structure can mitigate poor scaling with \textit{state} dimension, condensed QPs still scale adversely with input dimension as seen in \cite{Jerez:2011}. As embedding a non-autonomous system requires embedding both state \textit{and} control this puts both sparse- and condensed-form QPs at risk of computational intractability if MPC is approached in a conventional manner.

The proposed HAVOK approach is well-suited to resolving these issues via low-order approximations given its connection to the SVD, which is a staple of many existing model reduction algorithms (see e.g. \cite{Gugercin:2000}). Denoting an embedded state vector $z$ corresponding to an entire column of $H$, it is clear that:
 
 \begin{equation}\label{eq:Projection}
 	\begin{gathered}
 		 	z = USv^T \Leftrightarrow v^T = S^{-1}U^Tz 
 		 	\\ z \in \mathbb{R}^{n \cdot n_d \times 1} \wedge v^T \in \mathbb{R}^{n \cdot n_d \times 1}
 	\end{gathered}
 \end{equation}

where $v^T$ is the corresponding state vector in PCT coordinates. It is possible to define a standard convex MPC problem using this relationship:

 \begin{equation}\label{eq:MSMPCIntegrator}
	\begin{aligned}			
		\displaystyle{\min_{\eta, \Delta u}} \quad &\sum_{k=0}^{N_{p}} \eta(k)^TQ\eta(k)+\sum_{k=0}^{N_{u}-1} \Delta u_v(k)^TR\Delta u_v(k) +  \lVert \epsilon \rVert^2 \\				
		\text{subj. to} \quad &\eta(0) = \eta_0 				 \\
		&\Delta u_v(0) = \Delta u_{v_0} \\
		&\eta(k+1) = A_{\eta}\eta(k) + B_{\eta}\Delta u_v(k) 		\\
		&\Delta x_{\min} \leq C_cP_{\uparrow}\Delta v^T(k) \leq \Delta x_{\max} \\
		&\Delta u_{\min} \leq C_c P_{\uparrow} \Delta u_v(k) \leq \Delta u_{\max} \\ 		
		&x_{\min} \leq LC_cP_{\uparrow}\Delta v^T(k) + x_0 \leq x_{\max} \\
		&u_{\min} \leq LC_cP_{\uparrow}\Delta u_v(k) + u_0 \leq u_{\max} \\ 		
		&\eta(N_p)  \leq \epsilon
	\end{aligned}
\end{equation}

where $P_{\uparrow} = US$ as in \cref{eq:Projection}, $C_c$ picks out the elements of a delay-embedded vector corresponding to $d = 0$, $L$ is a lower triangular matrix that sums over the horizon, $\epsilon$ is a slack variable, and $\eta, A_\eta, B_\eta$ are the augmented state and velocity-form matrices that ensure integral action as per \cite{Pannocchia:2015}, i.e.:

\begin{equation}\label{eq:MSMPCAugmentedMatrices}
	A_{\eta} = \begin{bmatrix}
		A_v & 0 \\ C_cP_{\uparrow}A_v & I
	\end{bmatrix}, \ B_{\eta} = \begin{bmatrix}
		B_v \\ C_cP_{\uparrow}B_v
	\end{bmatrix}, \ \eta = \begin{bmatrix}
	\Delta v^T \\ x-\bar{x}
\end{bmatrix}
\end{equation}

where $\bar{x}$ is the output reference and $\Delta v^T(k) = v^T(k)-v^T(k-1)$. The initial conditions $\Delta z_{u_0}, \Delta v_0^T$ can be obtained by storing historical inputs and outputs as vectors, projecting the latter onto the HAVOK subspace, and performing finite differencing at each timestep.

Suppose now that we truncate the SVD to rank $r$ such that:

 \begin{equation}\label{eq:ProjectionTrunc}
	z \approx U_rS_rv_r^T,\quad z \in \mathbb{R}^{n \cdot n_d \times 1} \wedge v_r^T \in \mathbb{R}^{r \times 1}
\end{equation}

\cref{eq:ProjectionTrunc} is a rank-deficient system of equations under any amount of truncation when $v_r^T$ is taken as the independent variable. This implies that \cref{eq:MSMPCIntegrator} is not guaranteed to yield a causal sequence of inputs; indeed, we find it generally does not. We therefore propose a structure that \textit{can} guarantee a causal sequence while still exploiting the dimensionality reduction of HAVOK.

We first define the \textit{downshifting matrix} $D$ and $i$th \textit{mapping vector} $m_i$, which are structured such that for e.g. $n_d = N_p = 3$:

\begin{equation}\label{eq:AuxLinAlg}
	D = 
	\begin{bmatrix}
		0 & 0 & 0 \\
		1 & 0 & 0 \\
		0 & 1 & 0
	\end{bmatrix}, 
	\ m_1 = \begin{bmatrix} 1 \\ 0 \\ 0 \end{bmatrix}
	\ m_2 = \begin{bmatrix} 0 \\ 1 \\ 0 \end{bmatrix}
\end{equation}

Now, if the initial delay-embedded control input $z_{u_0}(k) = \begin{bmatrix} u(k-1) & u(k-2) & u(k-3) \end{bmatrix}^T$ is known, future embedded inputs can be constructed as a function of the predicted inputs $u(j|k)$ at time $k$, $D$, and the $m_i$ vectors. For example, the input vector $z_u(k+1|k)$ can be written as:

\begin{equation}\label{eq:EmbedConstructExample}
	\begin{aligned}
		z_u(k+1|k) &= m_1 u(k+1|k) + m_2 u(k|k) + D^2 z_{u_0}(k) \\
		& = \begin{bmatrix}
			u(k+1|k) \\ 0 \\ 0 
		\end{bmatrix} + 
		\begin{bmatrix}
			0 \\ u(k|k) \\ 0 
		\end{bmatrix} 
		+ 
		\begin{bmatrix}
			0 \\ 0 \\ u(k-1) 
		\end{bmatrix}
	\end{aligned}
\end{equation}

This extends trivially to the $\Delta u$ case and enforces causality by construction. Recall now that the dynamics in HAVOK coordinates are given by:

\begin{equation}\label{eq:HAVOKDyn}
		\eta(k+1) = A_\eta \eta(k) + B_\eta \Delta u_v(k)
\end{equation}

This can be rewritten as a function of $\Delta z_u(k)$ via the projection matrix $P_\downarrow = S_r^{-1}U_r^T$:

\begin{equation}\label{eq:DelayDyn}
	\eta(k+1) = A_\eta \eta(k) + B_\eta P_\downarrow \Delta z_u(k)
\end{equation}

Using the formalism from \cref{eq:EmbedConstructExample}, we can then restate \cref{eq:DelayDyn} in terms of $\Delta u(k)$ and $\Delta z_{u_0}(k)$:

\begin{equation}\label{eq:CausalDyn}
		\begin{aligned}
		\eta(k+1)=A_\eta\eta(k) &+ B_\eta P_\downarrow \sum_{i = 1}^{n_d} m_i \Delta u(k-i) \\ &+ B_\eta P_\downarrow D^{k} \Delta z_{u_0}(k)
		\end{aligned}
\end{equation} 

We can now reformulate \cref{eq:MSMPCIntegrator} in terms of the current input $\Delta u(k)$ and the initial embedded vector $\Delta z_{u_0}(k)$:

 \begin{equation}\label{eq:MSMPCIntegratorCausal}
	\begin{aligned}			
		\displaystyle{\min_{\eta, \Delta u}} \quad &\sum_{k=0}^{N_{p}} \eta(k)^TQ\eta(k)+\sum_{k=0}^{N_{u}-1} \Delta u(k)^TR\Delta u(k) + \lVert \epsilon \rVert^2 \\
		\text{subj. to} \quad &\eta(0) = \eta_0 				 \\
		&\Delta z_u(0) = \Delta z_{u_0} \\
		&\eta(k+1) = A_{\eta}\eta(k) +  B_\eta P_\downarrow \Delta z_u(k)	\\
		&\Delta x_{\min} \leq C_cP_{\uparrow}\Delta v(k) \leq \Delta x_{\max} \\
		&\Delta u_{\min} \leq \Delta u(k) \leq \Delta u_{\max} \\ 		
		&x_{\min} \leq LC_cP_{\uparrow}\Delta v(k) + x_0 \leq x_{\max} \\
		&u_{\min} \leq L\Delta u(k) + u_0 \leq u_{\max} \\ 		
		&\eta(N_p)  \leq \epsilon
	\end{aligned}
\end{equation}

This corresponds to the following sparse QP problem:

\begin{equation}\label{eq:QPProb}
	\begin{aligned}
		&\displaystyle{\min_{\eta, \Delta u}} \quad \frac{1}{2} \begin{bmatrix}\eta^T & \Delta u^T \end{bmatrix} \begin{bmatrix}Q_{\text{blk}} & 0 \\ 0 & R_{\text{blk}} \end{bmatrix} \begin{bmatrix}\eta \\ \Delta u \end{bmatrix} \\
		&\text{subj. to} \quad \begin{bmatrix} F_\eta & F_u \end{bmatrix} \begin{bmatrix} \eta \\ u \end{bmatrix} = b-F_0,  
	\end{aligned}
\end{equation}

where $Q_{\text{blk}}, R_{\text{blk}}$ are block-diagonal concatenations of the state and control costs, the equality constraints are structured as in \cref{eq:MSMPCEqConstraintMats}, and we also impose the inequality constraints in \cref{eq:MSMPCIntegratorCausal}.

\begin{equation}\label{eq:MSMPCEqConstraintMats}
	\begin{gathered}
		F_\eta = \begin{bmatrix} 
			-\mathcal{I} & 0 & 0 & \hdots & 0 \\ 
			A_\eta & -\mathcal{I} & 0 & \hdots & 0 \\
			0 & A_\eta & -\mathcal{I} & \hdots & 0 \\ 
			0 & \ddots & \ddots & \ddots & \vdots \\
			0 & \hdots & 0 & A_\eta & -\mathcal{I} 
		\end{bmatrix}, \\
		F_u = \begin{bmatrix} 
			0 & 0 & 0 & 0 & 0 \\ 
			B_\eta P_\downarrow m_1 & 0 & 0 & 0 &  0 \\
			B_\eta P_\downarrow m_2 & B_\eta P_\downarrow m_1 & 0 & 0 & 0 \\ 
			\vdots & \ddots & \ddots & 0 & 0 \\
			B_\eta P_\downarrow m_{n_d} & \ddots & \ddots & \ddots & 0 \\
			0 & B_\eta P_\downarrow m_{n_d} & B_\eta P_\downarrow m_{n_d-1} & \hdots & 0 
		\end{bmatrix}, \\
		F_0 = \begin{bmatrix} 
			0 \\ B_\eta P_\downarrow D \Delta z_{u_0} \\ B_\eta P_\downarrow D^2 \Delta z_{u_0} \\ \vdots \\ B_\eta P_\downarrow D^{N_p} \Delta z_{u_0} 
		\end{bmatrix}, \ b = \begin{bmatrix} -\eta_0 \\ 0 \\ 0 \\ \vdots \\ 0 \end{bmatrix}
	\end{gathered}
\end{equation}

We note that these equality constraints also indirectly impose causality on $\eta$.

\section{Laboratory Setup}\label{sec:DistrictHeating}
\begin{figure}[h!]
	\begin{center}
		\includegraphics[width=8.4cm]{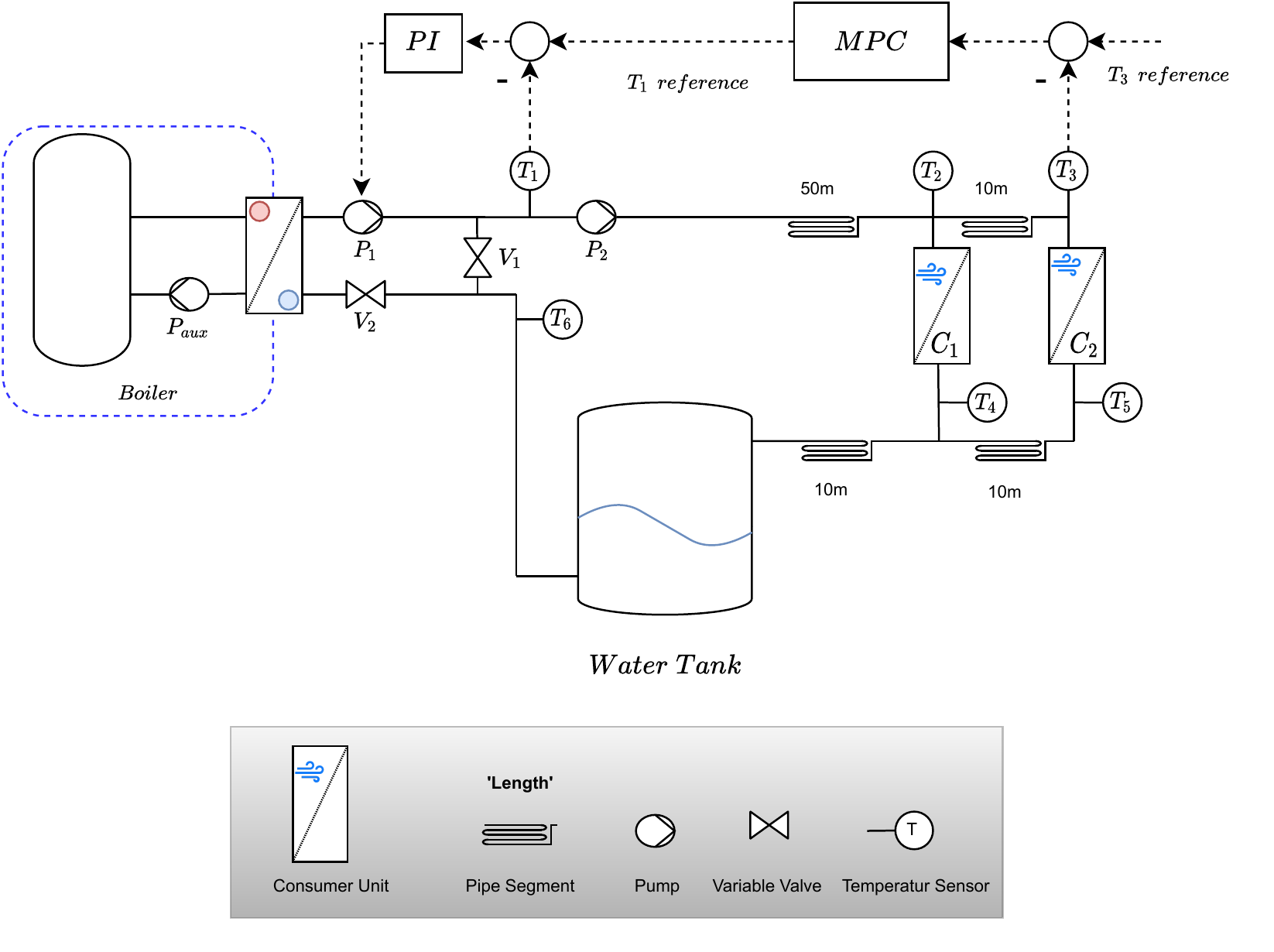}    
		\caption{Schematic of the AAU SWIL laboratory setup} 
		\label{fig:SWILSchematic}
	\end{center}
\end{figure}

\begin{figure}[h!]
	\begin{center}
		\includegraphics[width=8.4cm,height=4cm]{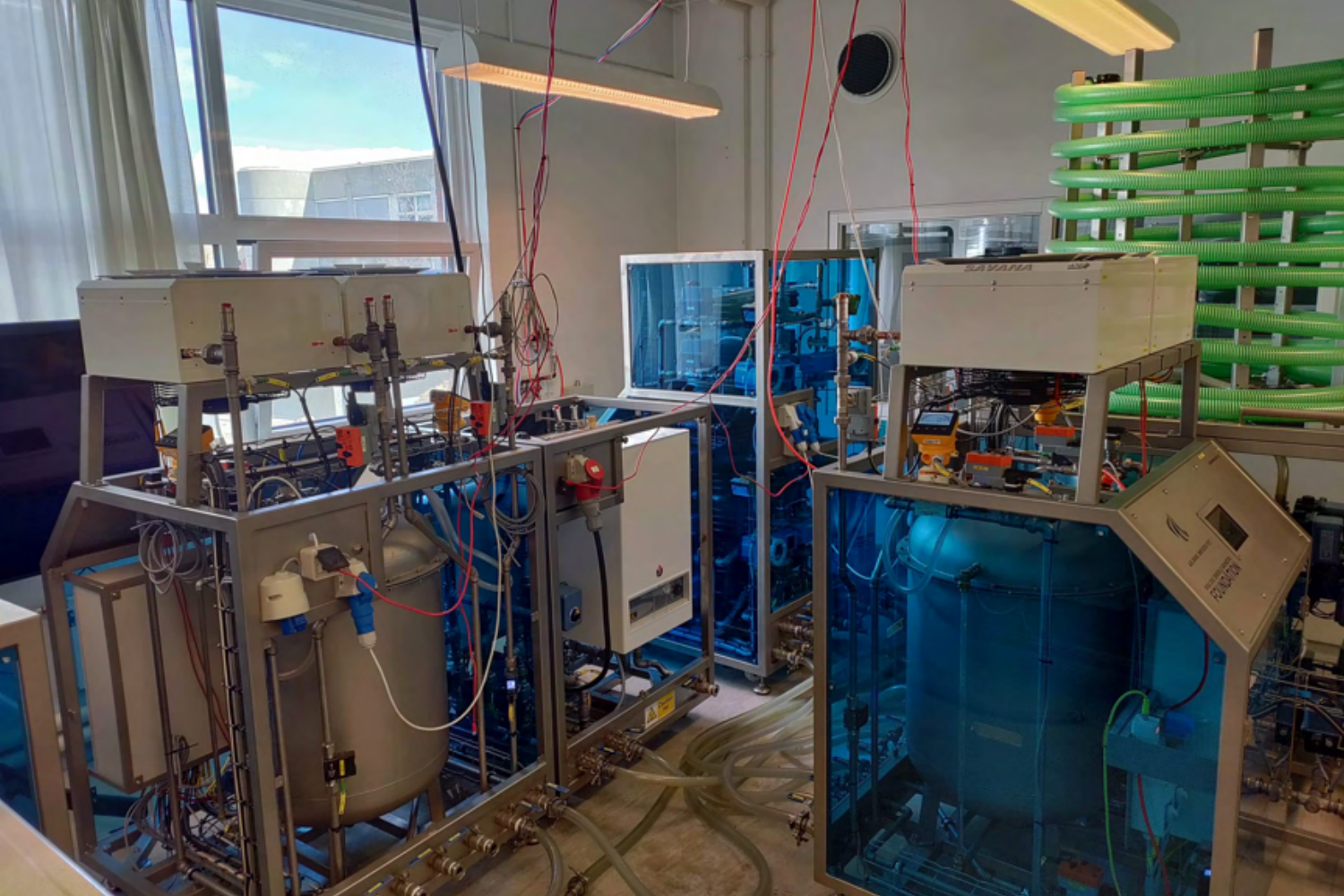}    
		\caption{Smart Water Infrastructure Laboratory (AAU-SWIL)} 
		\label{fig:SWILPicture}
	\end{center}
\end{figure}

The proposed HAVOK-MPC was tested in the AAU Smart Water Infrastructure Laboratory (AAU-SWIL). AAU-SWIL is a modular water laboratory comprising an assortment of units that can be arranged for laboratory-scale emulation of a wide variety of real-world systems (\cite{Val:2021}). The setup used to validate the performance of the proposed HAVOK-MPC is a single-supplier, dual-consumer DHS. The HAVOK-MPC acts as a central control unit (CCU) and generates the reference $\bar{T}_{in}$ to a local PI controller which controls the hot feedwater pumps in a mixing loop that supplies $T_{in}$. This is depicted schematically on \cref{fig:SWILSchematic} and photographically on \cref{fig:SWILPicture}.

The transport dynamics can be described by the advection-diffusion equation as given for the heat transport case in \cite{Bendtsen:2017}.

\begin{equation}\label{eq:Conv-diffPipe}
	A\rho c_p \frac{\partial\tau(l,t)}{\partial t}+v(t)A\rho c_p \frac{\partial \tau(l,t)}{\partial l}=2\gamma \pi r(T_{a}-\tau(l,t))
\end{equation}

where $\tau$ is the temperature of the fluid in the pipe, $A$ is the pipe cross-sectional area, $v(t) = q(t)/A$ is the mean flow velocity, $\rho$ is the density of the transport medium, $c_p$ is the fluid heat capacity, $T_a$ is the ambient temperature, and $l$ is length along the pipe measured from the supply point. If good insulation is assumed, heat losses through the pipe walls can be neglected, and \cref{eq:Conv-diffPipe} can be reduced to:

\begin{equation}\label{eq:Conv-diffPipeNoLoss}
	\frac{\partial \tau(l,t)}{\partial t}+\frac{q(t)}{A}\frac{\partial \tau(l,t)}{\partial l}=0
\end{equation}

As shown in \cite{Hansen:2011}, flow-dependent delay functions of the type:

\begin{equation}\label{eq:Conv-DiffFinal}
	\tau(l_i,t)=T_{in}(t-d_{i})  \quad \wedge \quad d_i = d_i(q_i) = \frac{\alpha_i}{q_i}
\end{equation}

are solutions to this transport equation. This structure of the solution, where the temperature distribution depends on past values of a measurable input, directly motivates the use of the delay coordinates that underpin the HAVOK modelling approach. Furthermore, the HAVOK approach has been shown in \cite{Brunton:2016} to inherently produce tridiagonal operators, which as demonstrated in \cite{Baddoo:2023} agrees with the local structure of a spatially discretized advection equation. 

In addition to heat transport, several other aspects of the system impact the overall dynamics. The supply heat $T_{in}$ is produced via a boiler and a mixing loop: 

\begin{equation}\label{eq:TempInModel}
	\dot{T}_{in}(t)=\frac{1}{V_{s}}q_{in}(t)(T_{out}(t)-T_{in}(t))+Q(t)
\end{equation}

where $Q(t)$ is the power provided by the boiler via the hot feedwater, and $T_{out}$ is the return temperature, which is a simple mixing of the outlet temperatures of each consumer:

\begin{equation}\label{eq:TempOutModel}
	T_{out}(t) = \frac{1}{q_{in}}\sum_{i=1}^{p}q_{i}T_{i}(t)
\end{equation}

where $p$ is the total number of consumers. The temperature at each consumer behaves according to the first-order differential equation:

\begin{equation}\label{eq:HeatExchangerModel_Energies}
	\dot{{T}_{i}}(t)= \frac{1}{V_i}q_{i}(t)(T_{in}(t-d_{i})-T_{i}(t))-w_{i}(t)
\end{equation} 

where $V_i$ is the effective volume of the $i$th heat exchanger and $w_{i}(t)$ is the load at the $i$th consumer. 

When taken altogether, \cref{eq:TempInModel,eq:TempOutModel,eq:HeatExchangerModel_Energies} suggest that the return flow has a significant impact on the supply temperature, and therefore on the dynamic response seen from supply temperature reference $\bar{T}_{in}$ to consumer temperatures. This implies that the consumer loads $w_i(t)$ (as well as any other unmeasured losses) constitute latent dynamics that must be captured by the HAVOK model to accurately predict the necessary supply temperature reference. Additionally, the HAVOK model must also capture the closed-loop dynamics of the local PI controller. While the boiler could be controlled directly by the HAVOK-MPC, this is not desirable as the timescale is significantly shorter than the transport delay, and this would necessitate very large delay embeddings. 

\section{Experimental Procedure and Results}\label{sec:LabResults}

\begin{figure}[h!]
	\begin{center}
		\includegraphics[width=8.4cm]{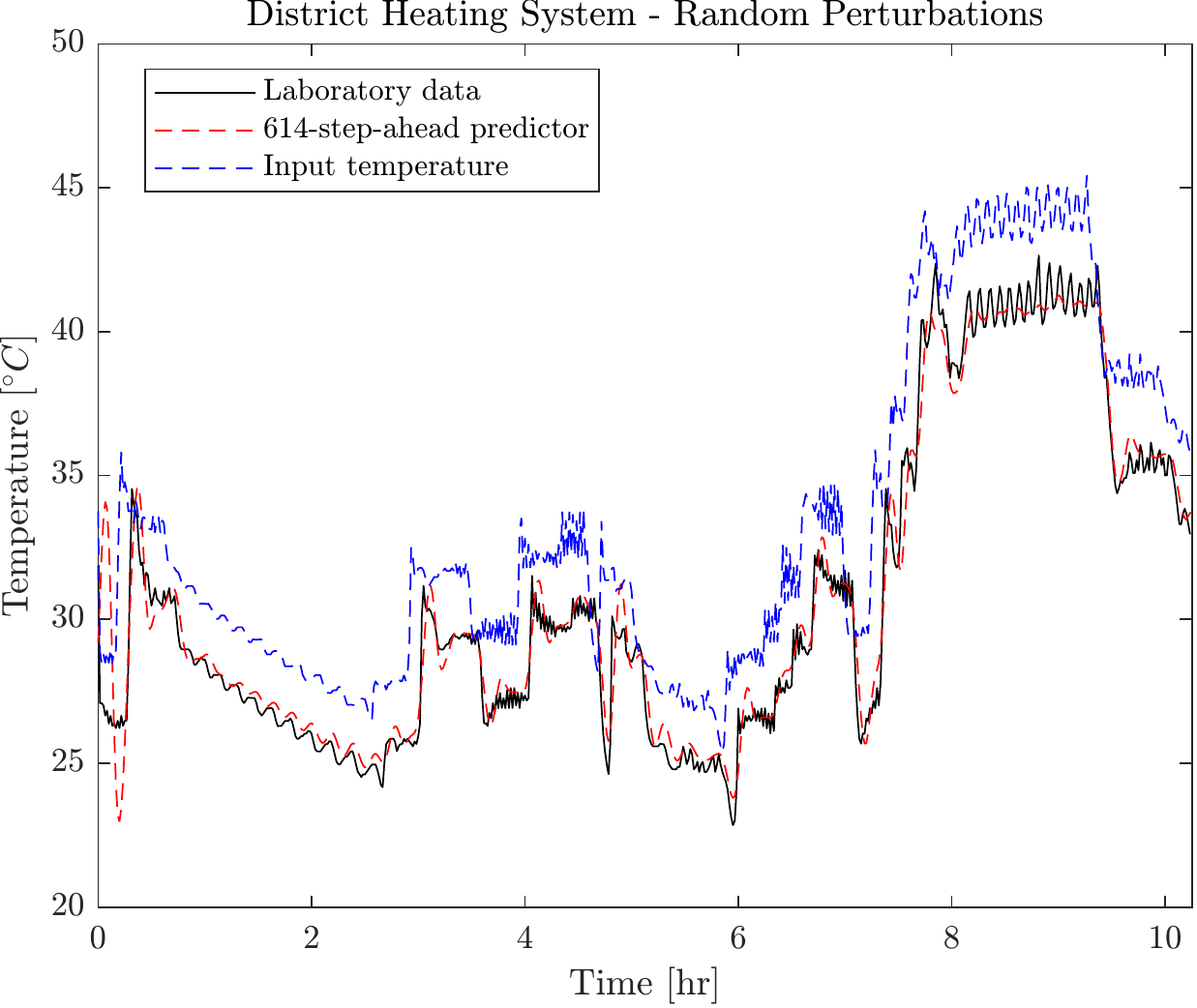}    
		\caption{HAVOK model predictive performance} 
		\label{fig:SWILPredictor}
	\end{center}
\end{figure}

The HAVOK model identification procedure is as follows: the mixing loop setpoint $\bar{T}_{in}$ is perturbed uniformly in time and value for approximately one day. Ambient temperature is kept near-constant using an adjacent HVAC laboratory, and flow out of the mixing loop is kept constant. A HAVOK model with $T_s = 60\text{s}$, $n_d = 20$, and $r = 5$ is constructed using the first $12$ hours of data, then evaluated against the remainder. We note that the embedding length is chosen exclusively based on a conservative upper bound (approximately $4$ times the true delay) on the system delay, while the odd number of modes enforces a model structure with one critically damped eigenvalue and two conjugate eigenpairs (\cite{Hirsh:2021}). Results can be seen on \cref{fig:SWILPredictor}. We remark that the predictive performance is excellent despite the long ($\approx 10$ hours corresponding to $614$ samples) horizon and unmeasured return flow dynamics, and that the low-order model structure captures the averaged temperature behaviour while neglecting spurious oscillatory elements of the mixing loop's closed-loop response.

\begin{figure}[h!]
	\begin{center}
		\includegraphics[width=8.4cm]{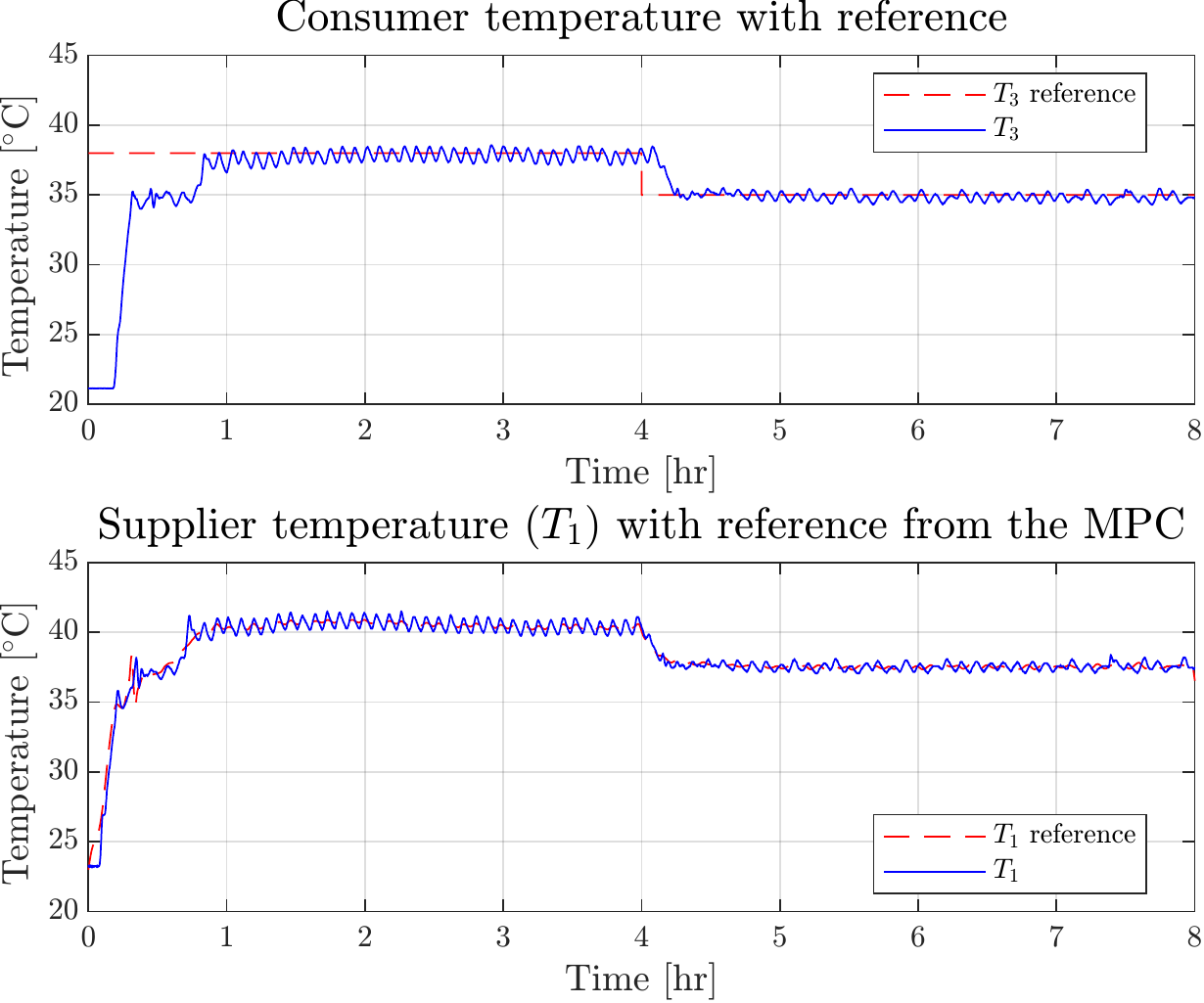}    
		\caption{HAVOK-MPC reference-tracking performance} 
		\label{fig:SWILResults}
	\end{center}
\end{figure}

The tracking performance of the proposed controller is then evaluated over an $8$-hour trajectory. We perform an initial step up from ambient, followed by a smaller step down after $4$ hours. As can be seen on \cref{fig:SWILResults}, the HAVOK-MPC smoothly adjusts the reference to the mixing loop $T_1$ in both cases, and stabilizes the supplier temperature at the desired setpoints within the span of an hour. No under- or overshoot is seen in either case despite the sizeable transport delay and temperature drop over the proximal consumer $C_1$, and the HAVOK-MPC stops adjusting $T_1$ at the correct value well before $T_3$ has settled, indicating that it correctly anticipates the transport delay. The small oscillations are clearly due to the mixing loop rather than the HAVOK-MPC, as the reference signal is smooth.

\section{Conclusion}\label{sec:Conclusion}
This paper has presented and practically validated a novel MPC formulation that exploits the HAVOK modelling framework. We have considered a single-supplier, dual-consumer DHS with a significant transport delay and temperature drop, as well as non-trivial latent dynamics, and verified the performance of the proposed HAVOK-MPC controller, which stabilizes consumer temperature at the desired values with minimal deviation in a laboratory setting. Furthermore, we have demonstrated that the proposed controller enforces a causal sequence of inputs while compressing the size of the MPC problem, and that the complexity induced by enforcing causality is bounded from above by the prediction horizon regardless of embedding length.

Critically, the proposed strategy is agnostic to the structure of the DHS and requires only knowledge of a rough upper bound on transport delay. It is also viable in the low-data limit, as demonstrated by the recovery of a good HAVOK model using fewer than $1000$ data points, and performant in spite of a local controller with unknown, non-negligible dynamics. This suggests its realistic applicability to distributed real-world systems with significant parameter uncertainty and limited data availability. We expect that the proposed framework can be extended to handle time-varying flow rates and consumer loads locally using an LPV approach such as the one described in \cite{Cisneros:2020}, or globally using the Bilinear DMD algorithm described in \cite{Goldschmidt:2021}.

Finally, while it is beyond the scope of this paper, future work should examine the performance of the demonstrated HAVOK-MPC relative to existing methods for control of time delay systems, particularly those that do not require any \textit{a priori} knowledge of the delay.

\begin{ack}
	The authors would like to acknowledge the help and advice of Saruch Satishkumar Rathore, particularly with the operation of the experimental setup.
\end{ack}

\bibliographystyle{harvard}
\bibliography{ifacconf}             

\end{document}